\documentclass[12pt,preprint]{emulateapj}
\shorttitle{Properties of AGN Selected by Optical Variability}
\shortauthors{T. Morokuma et al.}
\begin{document}
\title{
The Subaru/XMM-Newton Deep Survey (SXDS) - VI. 
Properties of Active Galactic Nuclei Selected by Optical Variability
\footnote{Based in part on data collected at Subaru Telescope, 
which is operated by the National Astronomical Observatory of Japan.}
}
\author{
Tomoki Morokuma\altaffilmark{1,2}, 
Mamoru Doi\altaffilmark{2}, 
Naoki Yasuda\altaffilmark{3}, 
Masayuki Akiyama\altaffilmark{4}, 
Kazuhiro Sekiguchi\altaffilmark{1,4}, 
Hisanori Furusawa\altaffilmark{4}, 
Yoshihiro Ueda\altaffilmark{5}, 
Tomonori Totani\altaffilmark{5}, 
Takeshi Oda\altaffilmark{5,6}, 
Tohru Nagao\altaffilmark{1,6}, 
Nobunari Kashikawa\altaffilmark{1}, 
Takashi Murayama\altaffilmark{7}, 
Masami Ouchi\altaffilmark{8,9}, 
Mike G. Watson\altaffilmark{10}
}
\altaffiltext{1}{Optical and Infrared Astronomy Division, National Astronomical Observatory of Japan, 2-21-1 Osawa, Mitaka, Tokyo 181-8588, Japan}
\altaffiltext{2}{Institude of Astronomy, Graduate School of Science, University of Tokyo, 2-21-1, Osawa, Mitaka, Tokyo 181-0015, Japan}
\altaffiltext{3}{Institute for Cosmic Ray Research, University of Tokyo, Kashiwa, Chiba 277-8582, Japan}
\altaffiltext{4}{Subaru Telescope, National Astronomical Observatory of Japan, 650 North A'ohoku Place, Hilo, HI 96720, USA}
\altaffiltext{5}{Department of Astronomy, Kyoto University, Sakyo-ku, Kyoto 606-8502, Japan}
\altaffiltext{6}{JSPS Fellow}
\altaffiltext{7}{Astronomical Institute, Graduate School of Science, Tohoku University, Aoba, Sendai 980-8578, Japan}
\altaffiltext{8}{Space Telescope Science Institute, 3700 San Martin Drive, Baltimore, MD 21218, USA}
\altaffiltext{9}{Hubble Fellow}
\altaffiltext{10}{Department of Physics and Astronomy, University of Leicester, Leicester LE1 7RH, UK}
\email{tmorokuma@optik.mtk.nao.ac.jp}

\begin{abstract}
We present the properties of active galactic nuclei (AGN) selected 
by optical variability in the Subaru/XMM-Newton Deep Field (SXDF). 
Based on the locations of variable components and light curves, 
$211$ optically variable AGN were reliably selected. 
We made three AGN samples; X-ray detected optically non-variable AGN (XA), 
X-ray detected optically variable AGN (XVA), 
and X-ray undetected optically variable AGN (VA). 
In the VA sample, we found a bimodal distribution of the ratio 
between the variable component flux and the host flux. 
One of these two components in the distribution, a class of AGN 
with a faint variable component $i'_{\rm{vari}}\sim25$ mag in 
bright host galaxies $i'\sim21$ mag, is not seen in the XVA sample. 
These AGN are expected to have low Eddington ratios if we naively 
consider a correlation between bulge luminosity and black hole mass. 
These galaxies have photometric redshifts $z_{\rm{photo}}\sim0.5$ and 
we infer that they are low-luminosity AGN with radiatively inefficient 
accretion flows (RIAFs). 
The properties of the XVA and VA objects and the differences from 
those of the XA objects can be explained within the unified scheme for AGN. 
Optical variability selection for AGN is an independent method and could 
provide a complementary AGN sample which even deep X-ray surveys have not found. 
\end{abstract}

\section{Introduction}\label{sec:intro}
Recent deep X-ray surveys have found many low-luminosity and obscured 
active galactic nuclei (AGN) and revealed luminosity-dependent cosmological 
evolution of AGN \citep*{ueda2003,barger2005}. 
The obscured fractions of AGN increase with decreasing X-ray luminosity 
\citep*{ueda2003,lafranca2005,akylas2006}. 
On the other hand, at optical wavelengths, many AGN surveys have been 
carried out by taking advantage of the blue optical colors of AGN, 
which are a common characteristic of unobscured (or type-1) AGN. 
However, the blue colors are difficult to recognize for AGN with dust 
obscuration and host galaxy contamination. 
Optical variability has been observed in almost all luminous AGN, 
i.e. quasars, on time scales of months to years 
\citep*{hook1994,giveon1999,devries2003,vandenberk2004,devries2005,sesar2006}. 
AGN selection by optical variability is less affected by host galaxy 
contamination than selection by blue optical color 
if the variable components can be extracted. 
Several SDSS results have showed that the optical variability of less 
luminous AGN is larger and this illustrates the usefulness of optical 
variability as a tracer of low-luminosity AGN. 
Variability studies using the Hubble Space Telescope (HST) actually found 
several tens of galaxies with variable nuclei down to $V,I,i'\sim27-28$ 
mag \citep*{sarajedini2000,sarajedini2003,sarajedini2006,cohen2006}. 
Although deep X-ray observations have been carried out with the Chandra 
and XMM-Newton satellites in the HST survey fields, 
there is a significant fraction ($>70\%$) of optically variable AGN without 
X-ray detection \citep*{sarajedini2006,cohen2006}. 
These authors showed that most of these X-ray non-detections can be explained 
in terms of small X-ray-to-optical flux ratios of the nuclear components. 
The number densities of variable AGN in their samples are comparable 
to those of X-ray detected AGN and these facts indicate that selection 
by optical variability is a powerful tool to find faint AGN populations 
which current deep X-ray observations may not be able to trace. 

There are also important results indicating the usefulness of optical 
variability as a tracer for AGN, especially for low-luminosity AGN. 
Radiatively inefficient accretion flows \citep[RIAFs;][]{quataert2001} 
are considered to have an accretion rate 
$\dot{m}(\equiv\dot{M}/\dot{M}_{\rm{Edd}})$ below a critical value 
in contrast with the standard disk model for luminous AGN. 
The spectral energy distributions of some nearby low-luminosity AGN 
have been explained in terms of RIAFs \citep*{chiaberge2006,nemmen2006}. 
\citet{totani2005} serendipitously found low-luminosity AGN 
in apparently normal bright galaxies at $z\sim0.3$ by optical variability 
in their cluster-cluster microlensing search using the images 
separated by several days and one month. 
This rapid and large fractional ($\sim100$\%) variability 
could be of blazar origin, but their emission line spectra 
and number densities support the RIAF interpretations. 
The low luminosities are also consistent with RIAFs. 
Their result indicated that the flare-ups of Sgr A$^\ast$ are not 
special phenomena and may be common in low-luminosity AGN 
in the distant universe. 
Multi-epoch ultraviolet images with HST revealed that most of the nearby 
low-ionization nuclear emission-line region (LINER) nuclei show significant 
variability with peak-to-peak amplitudes ranging from a few percent 
to 50\% \citep{maoz2005}. 
On the other hand, \citet{maoz2007} found that the properties of 
the SEDs of these LINERs and luminous AGN show continuous distributions, 
suggesting that thin accretion disks may persist to low luminosity. 

The optical continuum of AGN mainly comes from an accretion disk. 
The main origin of optical variability is still under debate; 
disk instability \citep*{rees1984,kawaguchi1998}, 
bursts of supernova explosions \citep{terlevich1992}, 
or microlensing \citep{hawkins1993}. 
However, if we assume that the optical variability of AGN 
also originates from an accretion disk, type-1 AGN should tend to 
show larger optical variability than type-2 AGN 
because we can directly see the accretion disk without it being 
obscured by a surrounding dust torus. 

In this paper, we investigate the X-ray, optical, and optical 
variability properties of faint variable AGN 
in the Subaru/XMM-Newton Deep Field (SXDF). 
The data was obtained by the Subaru/XMM-Newton Deep Survey (SXDS) project 
\citep{sekiguchi2004,sekiguchi2007}. 
\citet{morokuma2007} succeeded in constructing a statistical variable 
object sample and a well-classified AGN sample. 
We describe the AGN sample selections in \S\ref{sec:agnsample} and 
show the properties of optical-variability-selected AGN 
in \S\ref{sec:propva} and \S\ref{sec:type1agn}. 
We summarize our results in \S\ref{sec:summary}. 
In this paper, we use cosmological parameters of $\Omega_M=0.3, 
\Omega_\Lambda=0.7$, and Hubble constant $H_0=70$ km sec$^{-1}$ Mpc$^{-1}$. 
The AB magnitude system is used for optical photometry. 
We define $i'_{\rm{vari}}$ as the $i'$-band magnitude amplitude 
(minimum to maximum) of the variable components 
and $i'$ as the $i'$-band total magnitude. 

\section{AGN Sample}\label{sec:agnsample}
In this section, we describe our AGN sample selection. 
Our survey field, the SXDF, is a multi-wavelength project 
covering $\sim1.2$ deg$^2$. 
We use deep optical imaging data 
\citep*{furusawa2007,morokuma2007} 
taken with Suprime-Cam \citep{miyazaki2002} 
on the 8.2-m Subaru telescope for the optical variability investigation. 
X-ray imaging data with XMM-Newton satellite is also used 
for the AGN selection. 

\subsection{Optical Variability-Selected AGN Sample}\label{sec:variagnsample}
Our AGN sample selected by optical variability is based 
on the variable object sample constructed by \citet{morokuma2007}. 
By applying an image subtraction method \citep*{alard1998,alard2000} 
to multi-epoch (8-10 times from 2002 to 2005) $i'$-band 
deep ($i=25.2-26.8$ mag) imaging data obtained with Suprime-Cam, 
they found $1040$ variable objects among $\sim600,000$ objects, 
showing significant ($>5\sigma$) variability over $0.918$ deg$^2$. 
The detection limit for variable components 
is $i'_{\rm{vari}}\sim25.5$ mag, 
where $i_{\rm{vari}}$ is defined as the magnitude of differential flux 
between tha maximum and minumum. 
For almost all the variable objects, the host objects are unambiguous 
and their optical photometric properties such as magnitudes and colors 
cataloged in \citet{furusawa2007} are used. 
These authors classified non-stellar variable objects 
(including point sources with non-stellar colors) 
as AGN and supernovae (SNe) based on the
locations of the variable components within the host objects 
together with their light curves  
in the three pointings of Suprime-Cam 
($0.56$ deg$^2$, SXDF-C, SXDF-S, and SXDF-E) from 2002 to 2005. 
Well-classified variable AGN have variable components at their centers 
of the host objects (offsets between variable components and their 
host objects $<1.2$ pixel\footnote{Pixel scale of Suprime-Cam is 0\farcs202.}) 
and have non-SN-like light curves. 
Variable objects with these two properties are defined as 
in case 2 of \citet{morokuma2007}. 
The baselines of the light curves were not long or dense 
enough to discriminate AGN from SNe completely. 
There are many variable objects which have SN-like light curves 
and show variability lying at the centers of the host objects. 
These variable objects can be either SNe or AGN, and we do not 
include such objects in our variable AGN sample. 
Hence, we use a variability-selected AGN sample consisting of 
$211$ variable AGN in the region which overlaps the X-ray imaging field. 
We note that the number of case 2 objects ($228$) in \citet{morokuma2007} 
is slightly different from the number of variable AGN used in this paper 
because we focus on objects only within the X-ray imaging field. 

\subsection{X-ray-Selected AGN Sample}\label{sec:xrayagnsample}
In the SXDF, deep X-ray imaging observations were carried out 
with European Photon Imaging Camera (EPIC) on board XMM-Newton satellite. 
One deep ($\sim100$ ks) pointing and six shallower ($\sim50$ ks) pointings 
covered almost the entire Suprime-Cam field of the SXDF 
\citep*{ueda2007,akiyama2007}. 
The detection limit is $1\times10^{-15}$ erg$^{-1}$ cm$^{-2}$ s$^{-1}$ 
in the 0.5-2.0 keV band and $3\times10^{-15}$ erg$^{-1}$ cm$^{-2}$ s$^{-1}$ 
in the 2.0-10.0 keV band, respectively. 
The X-ray sources which we use in this paper have detection likelihood 
higher than nine in either energy band. 
The X-ray flux is calculated assuming a power-law X-ray spectrum 
with photon index $\Gamma=1.5$. 
In order to compare the properties of the X-ray selected AGN 
with those of the optical-variability-selected AGN, 
we use $327$ X-ray sources in the variability survey region 
where we selected $211$ optically variable AGN in \S\ref{sec:variagnsample}. 

\subsection{AGN Sample Classification}\label{sec:agnsampleclassification}
We classify these two, optical-variability-selected and X-ray-selected, 
AGN samples into three categories; 
1) X-ray detected, optically non-variable AGN ($238$ objects, hereafter ``XA''),
2) X-ray detected, optically variable AGN ($89$ objects, hereafter ``XVA''),
3) X-ray undetected, optically variable AGN ($122$ objects, hereafter ``VA''). 
Matching between the optically variable AGN and the X-ray detected AGN 
was done on the basis of the optical and X-ray positions, 
and the X-ray positional errors. 
We first assigned the nearest optical objects within $5\sigma$ of 
the X-ray positional errors from the X-ray centroids 
as the potential optical counterparts of the X-ray detected AGN. 
Then, we defined objects as ``XVA'' if the host objects of the optically 
variable objects are identical to the optical counterparts of the X-ray AGN. 

Spectroscopic redshifts were determined for $36$, $35$, and $9$ objects 
in the XA, XVA, and VA samples. 
Our spectroscopic observations were biased to X-ray detected objects 
and the number of VA objects with redshift determinations is small. 

For these three AGN samples, we calculated various statistical parameters 
such as the average, median, standard deviations, and 
the Kolmogorov-Smirnov (K-S) test probabilities between the samples. 
These values are summarized in Table \ref{tab:varixraycompare1} and 
Table \ref{tab:varixraycompare2}. 
Some of these values will be discussed in \S\ref{sec:type1agn}. 

\subsection{Variability Detection Completeness}\label{sec:complete_vari}
The detection efficiency of optical variability depends on 
not only the depth of the imaging data but also on observation time sampling. 
Variability detection itself depends on the depths of the images. 
All the Suprime-Cam images used in this paper have similar
depths ($i'\sim25.5-26.8$ mag) 
and we can detect object variability down to component amplitudes of 
$i'_{\rm{vari}}\sim25.5$ mag. 
The optical variability behavior of AGN differ from object to object 
and the detection completeness calculations for AGN are very complicated. 
\citet{morokuma2007} intensively examined the detection completeness 
and showed that the four-year baseline observations 
gave us an efficiency of $\sim80$\% at $i'\sim21$ mag and $\sim0$\% 
at $i'\sim24$ mag \citep[see \S5.2.2 in][]{morokuma2007}. 
We have Suprime-Cam observations with a four-year baseline from 2002 to 2005 
for all the objects which we use in this paper. 

\begin{figure}[htbp!]
\begin{center}
\epsscale{.8}
\includegraphics[angle=270,scale=0.35]{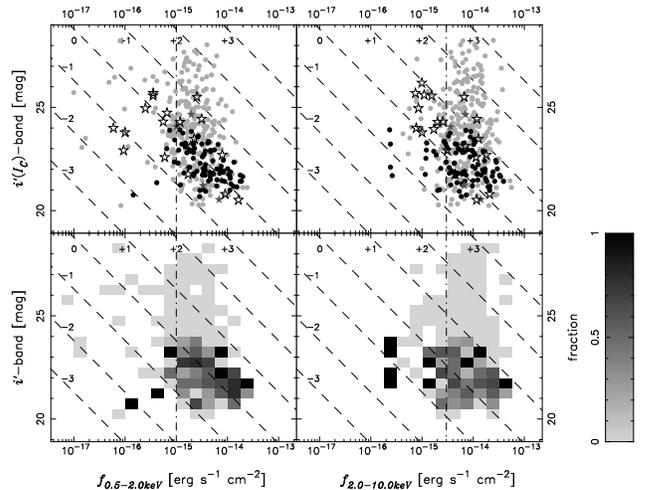}
\caption{
X-ray flux in the 0.5-2.0 keV band (left) and 2.0-10.0 keV band (right) 
and apparent $i'$-band magnitudes for the X-ray sources (XA objects, gray circles) 
and the X-ray detected optically variable objects (XVA objects, black circles) 
are shown in the upper panels. 
Detection limits in the 0.5-2.0 keV and 2.0-10.0 keV bands are 
$1\times10^{-15}$ erg$^{-1}$ cm$^{-2}$ s$^{-1}$ and 
$3\times10^{-15}$ erg$^{-1}$ cm$^{-2}$ s$^{-1}$ as indicated 
by the dot-dashed lines, respectively. 
Objects with X-ray flux below the detection limits are only plotted 
if detected above likelihood $9$ only in the other energy band. 
Optically variable AGN and non-variable AGN in the Groth Survey Strip 
\citep{sarajedini2006} are also plotted in dark gray stars and 
open stars, respectively. 
$I_C$-band magnitudes from \citet{vogt2005} are used for these objects, 
but we do not apply any band transformations. 
The fraction of X-ray sources whose optical variability is detected 
are also shown in gray scale in the lower panels. 
We note that zero fractions and regions where we have no X-ray sources 
are shown in light gray (not blank) and blank, respectively. 
The seven dashed lines represent constant X-ray-to-optical flux ratios of 
$\log(f_X/f_{i'})=+3,+2,+1,0,-1,-2,-3$ from top to bottom. 
\label{fig:xrayoptmag}}
\end{center}
\end{figure}

\begin{figure}[htbp!]
\begin{center}
\epsscale{.8}
\includegraphics[angle=270,scale=0.55]{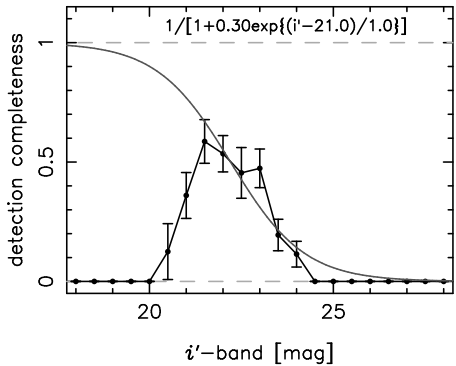}
\caption{
Filled circles show the detection completeness for optical variability 
among the X-ray sources in the SXDF. 
The cut-off at the bright end is caused by the exclusion of bright objects 
in our variability detection. 
The solid line is a function of the form 
$1/[1+a\times\exp\{(\rm{mag}-b)/c\}]$  fitted only in the range of 
$i'>21$ mag, 
because detection completeness in the brigher range could be affected 
by mis-subtraction or saturation. 
\label{fig:xrayoptmag4}}
\end{center}
\end{figure}

In the top panels of Figure \ref{fig:xrayoptmag}, 
we show the X-ray flux versus $i'$-band magnitude distributions 
for the XA and XVA objects. 
The fractions of X-ray sources showing optical variability are shown 
in the bottom panels of Figure \ref{fig:xrayoptmag} 
and Figure \ref{fig:xrayoptmag4} as a function of $i'$-band magnitude. 
These figures indicate that the variability detection efficiency 
for X-ray sources decreases down to zero at $i'=24-25$ mag 
and it is difficult to detect optical variability of 
X-ray sources with high X-ray-to-optical flux ratios. 
Figure \ref{fig:xrayoptmag} is further discussed in \S\ref{sec:type1agn}. 

In Figure \ref{fig:xrayoptmag}, 
we also plot the X-ray detected optically non-variable AGN ($25$ objects) 
and the X-ray detected optically variable AGN ($4$ objects) from 
one of similar studies \citep{sarajedini2006} for a comparison (these 
correspond to the XA and XVA objects in this paper). 
We plotted $I_C$-band magnitudes, which was available from 
\citet{vogt2005}, for the sample of \citet{sarajedini2006}. 
The difference of the bandpasses 
between $i'$-band and $I_C$-band is not large and 
we do not apply any band transformations. 
The observational properties of the objects plotted in this figure 
are derived from their Tables 6 and 7. 
The X-ray flux in each band is calculated using the full-band (0.5-10 keV) 
X-ray flux and hardness ratio 
(calculated as $f_{\rm{2.0-10.0keV}}/f_{\rm{0.5-2.0keV}}$) 
in their Table 6. 
The significance threshold for optical variability is set $3.2\sigma$, 
which is the same value as \citet{sarajedini2006} adopted. 
The differences in the distributions between our sample and 
their HST sample can be due to the differences of the depths 
of the observations. 


\section{Properties of AGN without X-ray detections}\label{sec:propva}
We first focus on the properties of the VA objects, which are 
defined as variable AGN without X-ray detections, 
and compare with those of the XA and XVA objects. 

\begin{figure}[htbp]
\begin{center}
\epsscale{.8}
\includegraphics[angle=270,scale=0.3975]{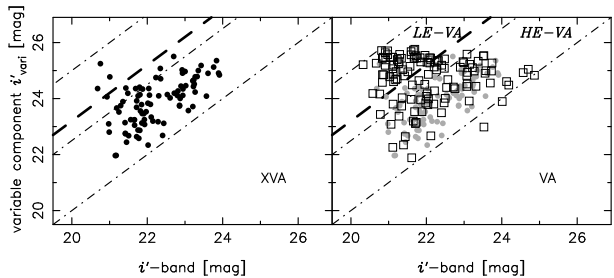}
\caption{
Distributions of $i'$-band magnitude versus variable component 
magnitude $i'_{\rm{vari}}$ of the XVA (filled circles, left panel) 
and VA (squares, right panel) objects. 
The XVA objects are also plotted in the right panel as 
gray filled circles for comparison. 
Dashed lines separating the LE-VA objects from HE-VA objects, 
$i'_{\rm{vari}}=1.0\times i+3.2$ 
(variable component flux is $0.05$ of the total flux), 
are indicated as thick dashed lines in both panels. 
The thin dot-dashed lines indicate constant ratios of 
variable components to total magnitudes of $0.01$, $0.1$, and $1$ 
(from left to right). 
\label{fig:agnvaricomp0}}
\end{center}
\end{figure}

\begin{figure}[htbp]
\begin{center}
\epsscale{.8}
\includegraphics[angle=270,scale=0.5300]{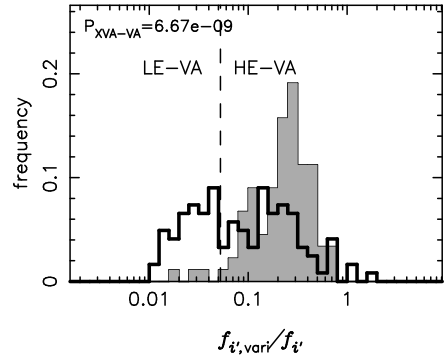}
\caption{
Distributions of the ratios between total flux and variable 
component flux of the XVA objects (gray histogram) and 
the VA objects (thick solid line histogram). 
The dashed line corresponds to the thick dashed line 
in Figure \ref{fig:agnvaricomp0}, 
separating the LE-VA and HE-VA objects. 
\label{fig:agnvaricomp7}}
\end{center}
\end{figure}

The distributions of the variable component magnitude 
$i'_{\rm{vari}}$ versus $i'$-band magnitude of the host objects 
for the XVA and VA objects are shown in Figure \ref{fig:agnvaricomp0}. 
Significant differences between the XVA and VA objects are seen. 
In the right panel of Figure \ref{fig:agnvaricomp0}, 
there are objects which have a faint variable component 
($i'_{\rm{vari}}\sim25$ mag) in bright galaxies ($i'\sim21$ mag), 
while there are only a few such objects seen in the distribution 
for the XVA objects. 
In addition, histograms of the ratios between variable component flux 
$f_{i',\rm{vari}}$ and total flux $f_{i'}$ shown 
in Figure \ref{fig:agnvaricomp7} marginally indicate a bimodal distribution 
suggesting that the VA objects may consist of two classes of AGN. 
The low K-S test probability ($6.67\rm{e}-09$, Table \ref{tab:varixraycompare1}) 
of the flux ratio distributions also indicates that 
these distributions are different. 
Accordingly we separate the VA sample into two classes by a dashed line, 
$i'_{\rm{vari}}=1.0\times i+3.2$ ($f_{i',\rm{vari}}=0.05\times f_{i'}$), 
in Figure \ref{fig:agnvaricomp0}; 
HE-VA objects ($73$ objects, below the line) and 
LE-VA objects ($49$ objects, above the line). 
Assuming that AGN optical variability (differential) flux , 
not amplitude, is roughly proportional to optical luminosity of the AGN
\footnote
{
AGN optical variability amplitude is larger for less luminous AGN 
\citep{vandenberk2004}, but AGN variability flux, which is defined as 
differential flux among observational epochs, is larger for more 
lumionus AGN because variability amplitude dependence on AGN luminosity 
is not large. 
}
, 
AGN with faint variable components are 
considered to be faint AGN. 
Given the correlation between supermassive black hole mass 
and bulge luminosity \citep{wandel1999}, 
AGN with larger ratios between the variable component flux and 
total flux can be naively interpreted as AGN with higher Eddington ratios. 
Thus it is expected that LE-VA objects have low Eddington ratios 
while HE-VA objects have high Eddington ratios. 
The LE-VA sample produces the difference of the distributions 
between the XVA and VA objects in Figure \ref{fig:agnvaricomp0}. 
This difference should not be due to any selection effects 
because the selection cuts are along horizontal and vertical 
directions in this figure. 

\begin{figure}[htbp]
\begin{center}
\epsscale{.8}
\includegraphics[angle=270,scale=0.336]{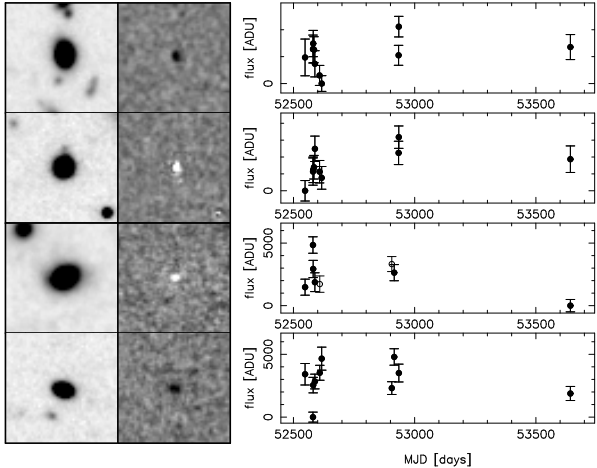}
\caption{
Examples images and light curves of four LE-VA objects 
with faint variable components, 
$i'_{\rm{vari}}\sim25$ mag, in bright galaxies. 
The left column shows the reference images before subtractions. 
The variable components in the subtracted images are seen 
in the right column images. 
Unreliable photometric points are plotted as open circles. 
\label{fig:vaobjimage}}
\end{center}
\end{figure}

The LE-VA objects are AGN with faint variable components in bright galaxies. 
These objects are similar to low-luminosity AGN in 
bright elliptical galaxies which were found using 
optical variability on time scales of several days to 
a month by \citet{totani2005}. 
\citet{totani2005} indicated that the rapid variability 
may be due to flare-ups in RIAFs rather than a blazar origin 
and noted the similarity to near-infrared flares of 
Sgr A$^\ast$ \citep{yuan2004}. 
RIAF disks have low accretion rates and low Eddington ratios, 
and tend to show flare-ups on short time scales. 
We show four examples of Suprime-Cam images and light curves 
of these objects in Figure \ref{fig:vaobjimage}. 
These objects are randomly selected from the LE-VA sample. 
Some light curves are likely to be those of flare-ups. 
If the LE-VA objects are really equivalent to AGN showing rapid 
variability as found by \citet{totani2005}, 
their variation time scales are expected to be shorter than 
those of the HE-VA objects on average. 
However, it is difficult to investigate the time scales of 
variability quantitatively because of the sparse time sampling. 
We tried evaluating two kinds of variability time scales: 
as the minimum time interval over which objects show significant 
($>5\sigma$) variability, and as the interval between maxima and minima. 
There are no significant differences for either time scale
between the LE-VA and HE-VA objects. 
It is not clear which objects show variability on shorter time scales. 
However, this does not reject the RIAF interpretation for LE-VA objects. 

\begin{figure}[htbp]
\begin{center}
\epsscale{.8}
\includegraphics[angle=270,scale=0.33656]{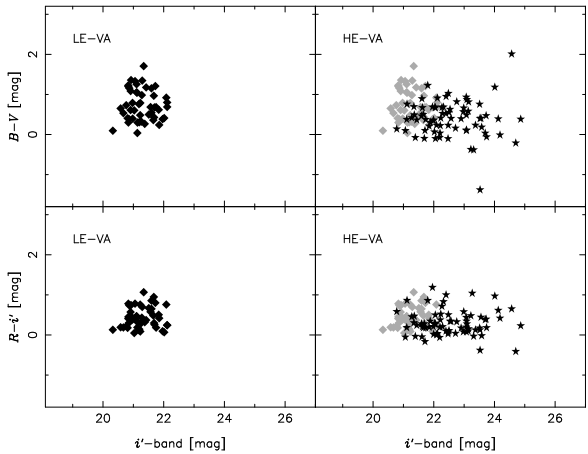}
\caption{
Optical color-magnitude diagrams of the LE-VA (filled squares, left column) 
and HE-VA (filled stars, right column) objects. 
The LE-VA objects are also plotted in the right column 
as gray filled squares for a comparison. 
\label{fig:agnoptcolor4}}
\end{center}
\end{figure}

Figure \ref{fig:agnoptcolor4} shows the optical color-magnitude 
distributions for the LE-VA and HE-VA objects. 
The K-S test probabilities for these distributions and 
their averages are given in Table \ref{tab:varixraycompare5} 
and Table \ref{tab:varixraycompare4}, respectively. 
The LE-VA objects have significantly redder $B-V$ colors 
than the HE-VA objects on average. 
In our sample, there are only a few objects which are selected 
as $B$-dropout objects. 
The intrinsically blue colors of AGN should remain blue 
in the observed $B-V$ colors even when redshifted. 
The red colors of LE-VA can be explained by large contamination 
by red host galaxies and might indicate that most of them are 
early-type galaxies at relatively low redshift. 
When we calculate the photometric redshifts for these galaxies 
without considering any AGN light contribution, 
the optimal spectral templates and redshifts are 
early-type galaxies at $z_{\rm{photo}}\sim0.5$ for most of the LE-VA objects, 
also supporting a low luminosity for these AGN. 

\begin{figure}[htbp]
\begin{center}
\epsscale{.8}
\includegraphics[angle=270,scale=0.33656]{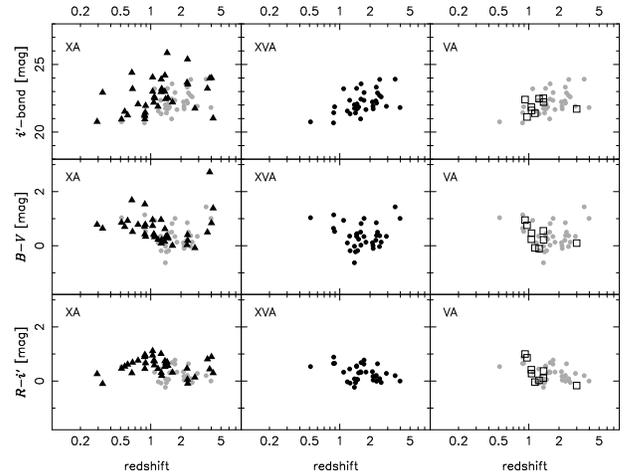}
\caption{
Redshift distributions of $i'$-band magnitudes and optical colors of 
the XA (filled triangles, left column), XVA (filled circles, central column), 
and VA (squares, right column) objects. 
The XVA objects are also plotted in the left and right columns 
in gray filled circles for comparison. 
\label{fig:agnredshift0}}
\end{center}
\end{figure}

The HE-VA objects also show a similar distribution to 
the XVA objects in Figure \ref{fig:agnvaricomp0}. 
Figure \ref{fig:agnredshift0} shows the redshift distributions 
for the XA, XVA, and VA objects. 
We have no spectroscopic identifications for the LE-VA objects 
and all the VA objects plotted in this figure belong to the HE-VA subsample. 
Most of the spectroscopically identified AGN in the XVA and VA samples 
are at high redshift ($z>1$) and the HE-VA objects are expected 
to be similar objects to the XVA objects. 
We interpret the X-ray non-detections of the HE-VA objects 
as deriving from the intrinsically wide distributions of 
X-ray-to-optical flux ratios of AGN \citep[e.g.,][]{anderson2007}, 
as seen in Figure \ref{fig:xrayoptmag}. 
If we assume that the X-ray-to-optical flux ratio distributions of the
optically variable AGN are independent of their brightness 
and the distributions for bright ($i'\sim22$ mag) XVA objects are the same 
as those for fainter XVA objects, 
there should be $\sim20$ VA objects just below the X-ray detection limit. 
The number of HE-VA objects is $73$, much larger than this estimate. 
However, many VA objects are as bright as $i'\sim21-22$ mag 
and the X-ray-to-optical flux ratio distributions of our XVA sample 
may not represent the entire intrinsic distributions 
even in the bright magnitude range. 
There can be AGN with lower X-ray-to-optical flux ratios for 
which we can detect their optical variability but cannot detect 
their X-ray emission. 

Thus we infer that the VA sample consists of two classes: 
low-luminosity AGN at relatively low redshift (LE-VA) 
and luminous AGN at high redshift (HE-VA). 
Other similar studies of optical variability-selected AGN 
with HST found that significant fractions ($\sim70\%$) of variable AGN 
in their samples were not detected in deep X-ray imaging with the Chandra 
or XMM-Newton satellites \citep*{sarajedini2006,cohen2006}. 
Our results, as well as HST results, indicate 
that optical variability can trace AGN classes 
which are not detected in deep X-ray surveys. 

\section{Are Optical-Variability-Selected AGN type-1?}\label{sec:type1agn}
As discussed in \S\ref{sec:intro}, it is natural to expect that 
objects showing optical variability are type-1 AGN because optical variability 
of AGN is considered to originate in their accretion disks. 

\begin{figure}[htbp]
\begin{center}
\epsscale{.8}
\includegraphics[angle=270,scale=0.33655]{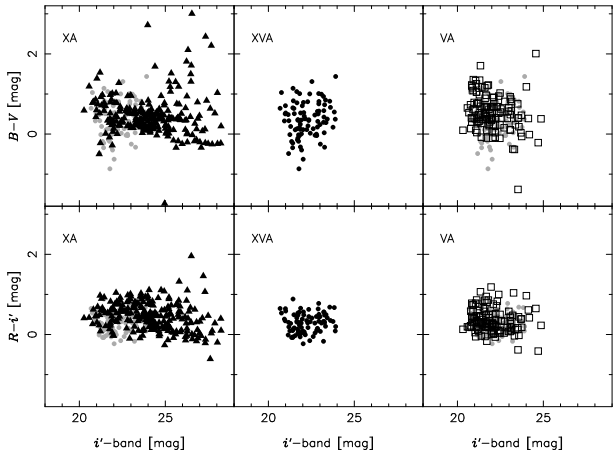}
\caption{
Optical color-magnitude diagrams for the XA, XVA, and VA objects. 
Symbols used are the same as those in Figure \ref{fig:agnredshift0}. 
\label{fig:agnoptcolor1}}
\end{center}
\end{figure}

We first compare the optical properties (magnitudes and colors) of 
the XA, XVA, and VA objects. 
Figure \ref{fig:agnoptcolor1} 
shows the distributions of $B-V$ and $R-i$ colors, and $i'$-band magnitude. 
Figure \ref{fig:agnoptcolor1}, as well as Figure \ref{fig:xrayoptmag}, 
clearly indicates that optical variability can be detected only 
for relatively brighter AGN ($i'<23.9$ mag) 
amongst X-ray detected AGN because of our variability detection limit. 
The distributions of only the XA sample go down to fainter magnitudes. 
The K-S test probabilities indicates that 
significant color differences are seen for red ($R-i'$ and $i'-z'$) 
colors in the observed frame while distributions of $B-V$ and $V-R$ 
colors are not different. 
However, the redshift distribution of the XA objects is different 
from those of the XVA and VA objects 
(Figure \ref{fig:agnredshift0}) and the differences of observed colors 
should be affected by the redshift distribution differences. 

\begin{figure}[htbp]
\begin{center}
\epsscale{.8}
\includegraphics[angle=270,scale=0.375]{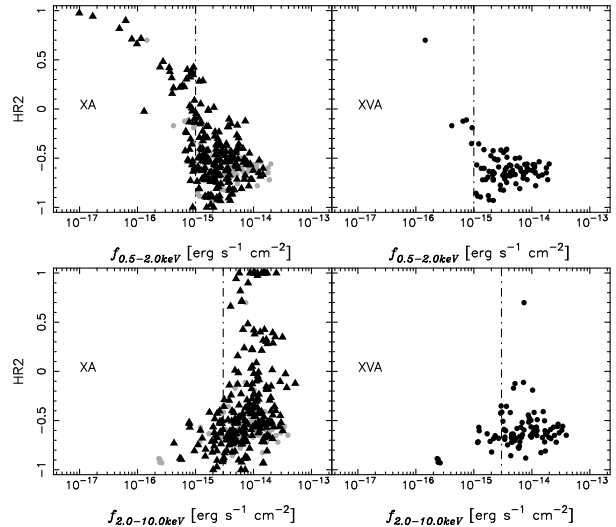}
\caption{
X-ray flux versus hardness ratio HR2 distributions of 
the XA and XVA objects. 
Symbols used are the same as those in Figure \ref{fig:agnredshift0}. 
Detection limits are shown as dot-dashed lines. 
Objects with X-ray flux below the detection limits are 
only plotted if detected above likelihood $9$ 
only in the other energy band. 
\label{fig:agnxrayprop0}}
\end{center}
\end{figure}

We now focus on the X-ray hardness ratio distributions. 
We define the hardness ratio, HR2, as the ratio of count rates 
in the 0.5-2.0 keV and 2.0-4.5 keV bands; 
$\rm{HR2}$$\equiv(H-S)/(H+S)$ 
($H$: count rate in the 2.0-4.5 keV band, 
$S$: count rate in the 0.5-2.0 keV band). 
By definition, HR2 can have values of $-1\leq\rm{HR2}\leq1$ and 
obscured, type-2, populations tend to have larger HR2 values 
because photons with higher energy can penetrate 
through the obscuring torus more efficiently. 
There may be a good correlation between AGN classification 
(type-1 or type-2) in X-rays and that deduced from optical 
spectroscopy \citep{ueda2003}. 
\citet{barger2005} showed that broad-line AGN with emission 
line widths above $2000$ km s$^{-1}$ are soft X-ray sources, 
while AGN with emission lines below this width have a wide 
range of X-ray colors. 
The correlation between optical obscuration and X-ray obscuration 
may be biased because classification using optical spectra requires 
good signal-to-noise ratios, 
but the hardness ratio can be a good parameter for evaluating 
optical obscuration. 
Figure \ref{fig:agnxrayprop0} shows the HR2 versus X-ray flux 
distributions for the XA and XVA objects. 
The HR2 distributions are significantly different. 
The XA objects tend to have higher HR2 values while the HR2 values 
of the XVA objects concentrate around $-0.6$. 
This can be naturally understood by considering the unified scheme 
of AGN because unobscured populations, in which we can see 
the nuclei directly, should show larger optical variability. 

\begin{figure}[htbp]
\begin{center}
\epsscale{.8}
\includegraphics[angle=270,scale=0.33656]{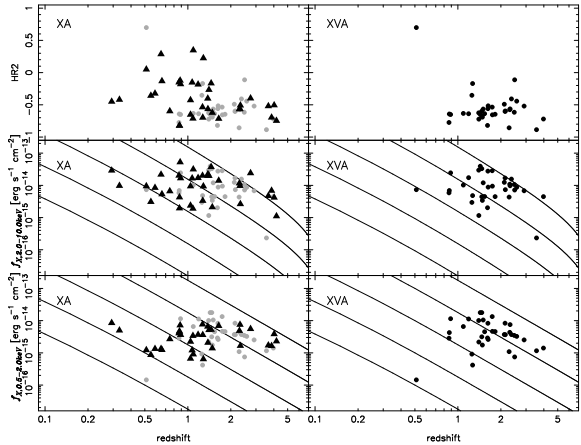}
\caption{
X-ray flux and hardness ratio HR2 as a function of redshift 
for the XA and XVA samples. 
Symbols used are the same as those in Figure \ref{fig:agnredshift0}. 
Solid lines indicate constant X-ray luminosity 
$L_{0.5-2.0\rm{keV}}$, $L_{2.0-10.0\rm{keV}}=
10^{41}, 10^{42}, 10^{43}, 10^{44}$, 
and $10^{45}$ erg s$^{-1}$ from bottom to top. 
\label{fig:agnredshift1}}
\end{center}
\end{figure}

The variability detection completeness also shows 
the differences of the selection effects 
between optical variability and X-ray detection.
Figure \ref{fig:xrayoptmag} indicates that the XVA objects 
(black circles) tend to have higher X-ray flux than 
the XA objects (gray circles). 
When we limit XA objects to those with $i'<23.9$ mag, 
which is the $i'$-band magnitude of the faintest XVA object, 
this tendency becomes weaker but still exists. 
High X-ray-to-optical flux ratios can be attributed to both 
optical faintness and large X-ray flux. 
Objects with extremely high X-ray-to-optical flux ratios 
are candidates for highly obscured luminous AGN, 
objects whose optical variability is more difficult 
to detect than unobscured AGN. 
The decline of the detection completeness for variability 
towards fainter magnitudes also contributes to this tendency, 
as well as the inclusion of obscured populations in the XA sample. 
The distributions of the hardness ratio HR2 and X-ray flux 
as a function of redshift shown in Figure \ref{fig:agnredshift1} 
also indicate that the XVA objects have lower hardness ratios 
and higher soft X-ray fluxes on average at any redshift. 

Lines of constant X-ray luminosity are shown in Figure \ref{fig:agnredshift1} 
assuming that the X-ray spectrum is well represented 
by a power-law with photon index $\Gamma=1.5$. 
\citet{ueda2003} showed that the fraction of X-ray 
type-2 AGN decreases with X-ray luminosity; 
this was also indicated in later studies \citep*{lafranca2005,akylas2006}. 
\citet{ueda2003} also found a possible similar effect 
in that the fraction of optical type-2 AGN increases 
with deceasing of X-ray luminosity although 
spectroscopic observational biases can affect 
this tendency because the host galaxy contaminations 
make it difficult to detect broad lines of AGN origin. 
Almost all of the XVA objects have X-ray luminosity higher than 
$\sim10^{43}$ erg s$^{-1}$ cm$^{-2}$, below which 
optical type-2 fraction of X-ray sources increases 
up to $0.4-1.0$ \citep{ueda2003}. 
The non-detections of optical variability 
for low-$z$ bright XA objects can be understood 
if they are obscured and low-luminosity populations. 

Although spectroscopic redshifts are available for 
only part of our AGN sample, as described 
in \S\ref{sec:agnsampleclassification}, 
the redshift distribution of XVA objects is biased towards 
slightly higher values than that of the XA objects, as is shown 
Figure \ref{fig:agnredshift0}. 
The median redshifts are $<z_{\rm{XA}}>=1.18$, $<z_{\rm{XVA}}>=1.48$,  
$<z_{\rm{VA}}>=1.40$, respectively. 
There are not many low-$z$ ($z<1$) XVA objects 
while there are many XA objects at such redshifts. 
The non-detections of optical variability from such 
bright XA objects can be explained if many of them 
are type-2 AGN with lower X-ray luminosities, 
less than $\sim10^{43}$ erg s$^{-1}$ cm$^{-2}$. 

The optical and X-ray properties of AGN can be summarized as follows. 
Compared with the XA objects, the XVA objects have lower HR2 values, 
smaller X-ray-to-optical flux ratios, higher X-ray flux, 
and appear at higher redshifts. 
These differences can be explained within the unified scheme of AGN 
considering the anti-correlation between luminosity and obscured fractions. 
We conclude that most of the optical-variability-selected AGN are type-1. 

\section{Summary}\label{sec:summary}
We investigated the X-ray, optical, and optical variability properties 
of X-ray-selected and optical-variability-seleted AGN samples in the SXDF. 
Amongst the VA objects, we found a class of AGN (LE-VA) with a faint 
variable component $i'_{\rm{vari}}\sim25$ mag in bright host galaxies 
$i'\sim21$ mag. 
In our definition the variability flux of these AGN are less than 
$0.05$ of their total flux, including host galaxy components. 
Our limited time sampling prevented us from determining the typical 
time scale of variability, but some of them show plausible flare-ups. 
They are similar to the low-luminosity AGN which \citet{totani2005} found. 
Therefore, we infer that they are low-luminosity AGN with RIAF at low redshift. 
The photometric redshifts, $z_{\rm{photo}}\sim0.5$, and extended 
morphologies of the LE-VA objects supports the idea that these AGN 
are low-luminosity objects. 
These low-luminosity AGN candidates may be similar to Sgr A$^\ast$ 
and some of nearby Seyfert nuclei, whose properties can be described 
in terms of RIAF. 

The XVA objects have lower X-ray hardness ratios than the XA objects on average. 
For the spectroscopically identified objects, XVA objects also have 
higher X-ray luminosity than the XA objects. 
These properties are consistent with those expected from the unified 
scheme for AGN and dependence of obscured fraction on X-ray luminosity. 
The XVA and VA objects are mainly unobscured, type-1 AGN. 

Although X-ray observations can effectively trace even obscured 
populations of AGN, optical variability selection for AGN is a useful 
method which is independent of X-ray selection 
and could provide a new AGN sample which even deep X-ray surveys have not found. 


\acknowledgements
This work was supported in part with a scientific research grant (15204012) 
from the Ministry of Education, Science, Culture, and Sports of Japan (MEXT).
M.A. is supported by a Grant-in-Aid for Young Scientists (B) 
from JSPS (18740118). 
This work is also supported in part with a scientific research grant (18072003) from the MEXT. 
We appreciate useful comments by Kimiaki Kawara and \v{Z}eljko Ivezi\'{c}. 
We are grateful to all members of the SXDS project. 
We also thank the anonymous referee for useful comments. 

\begin{deluxetable}{cccccc}
\tablewidth{0pt}
\tablecaption{K-S Test Probabilities\label{tab:varixraycompare1}}
\tablehead{
\colhead{sample1} &
\colhead{XA} &
\colhead{XA$_{\rm{bright}}$\tablenotemark{a}} &
\colhead{XA} &
\colhead{XA$_{\rm{bright}}$\tablenotemark{a}} &
\colhead{XVA}\\
\colhead{sample2} &
\colhead{XVA} &
\colhead{XVA} &
\colhead{VA} &
\colhead{VA} &
\colhead{VA}
}
\startdata
	redshift            & $1.33\rm{e}-03$  & $5.61\rm{e}-04$ & $3.30\rm{e}-01$ & $2.15\rm{e}-01$ & $2.55\rm{e}-02$\\
	$B-V$               & $3.20\rm{e}-01$  & $1.56\rm{e}-02$ & $1.27\rm{e}-02$ & $7.47\rm{e}-02$ & $1.47\rm{e}-02$\\
	$V-R$               & $3.91\rm{e}-01$  & $5.28\rm{e}-03$ & $4.97\rm{e}-01$ & $1.59\rm{e}-01$ & $2.16\rm{e}-01$\\
	$R-i'$              & $5.78\rm{e}-05$  & $1.27\rm{e}-07$ & $4.10\rm{e}-04$ & $7.96\rm{e}-07$ & $3.25\rm{e}-01$\\
	$i'-z'$             & $1.15\rm{e}-04$  & $9.87\rm{e}-07$ & $3.90\rm{e}-07$ & $4.21\rm{e}-10$ & $6.04\rm{e}-01$\\
	$i'$-band magnitude & $2.44\rm{e}-20$  & $4.12\rm{e}-02$ & $8.84\rm{e}-23$ & $2.30\rm{e}-04$ & $4.60\rm{e}-02$\\
	HR2                                & $1.07\rm{e}-07$  & $7.93\rm{e}-06$ & -          & -          & -       \\
	$\log(f_{X,\rm{0.5-2.0keV}})$      & $1.40\rm{e}-06$  & $3.27\rm{e}-03$ & -          & -          & -       \\
	$\log(f_{X,\rm{2.0-10.0keV}})$     & $8.74\rm{e}-01$  & $8.58\rm{e}-01$ & -          & -          & -       \\
	$\log(f_{X,\rm{0.5-2.0keV}}/f_{i'})$  & $2.80\rm{e}-10$  & $2.83\rm{e}-02$ & -          & -          & -       \\
	$\log(f_{X,\rm{2.0-10.0keV}}/f_{i'})$ & $9.36\rm{e}-10$  & $1.77\rm{e}-02$ & -          & -          & -       \\
	variable component $i'_{\rm{vari}}$ & -           & -          & -          & -          & $8.41\rm{e}-10$\\
	$\log(f_{i',{\rm{vari}}}/f_{i'})$   & -           & -          & -          & -          & $6.67\rm{e}-09$\\
\enddata
\tablecomments{K-S test probabilities between the sample 1 and sample 2 for the parameters.}
\tablenotetext{a}{The XA$_{\rm{bright}}$ sample consists of $199$ XA objects with $i'<23.9$ mag.}
\label{tab:varixraycompare1}
\end{deluxetable}
\begin{deluxetable}{cccc}
\tabletypesize{\scriptsize}
\tablewidth{0pt}
\tablecaption{Averages (Medians) and Standard Deviations of Parameters \label{tab:varixraycompare2}}
\tablehead{
\colhead{parameter} &
\colhead{XA} &
\colhead{XVA} &
\colhead{VA}
}
\startdata
	redshift      & $1.479\pm1.027\ (1.180)$ & $1.815\pm0.731\ (1.623)$ & $1.358\pm0.601\ (1.152)$\\
	              & $1.328\pm0.903\ (1.086)$ &                         &                        \\
	$B-V$         & $0.475\pm0.560\ (0.387)$ & $0.360\pm0.438\ (0.357)$ & $0.535\pm0.453\ (0.499)$\\
	              & $0.469\pm0.335\ (0.428)$ &                         &                        \\
	$V-R$         & $0.375\pm0.414\ (0.327)$ & $0.307\pm0.354\ (0.329)$ & $0.431\pm0.419\ (0.379)$\\
	              & $0.509\pm0.325\ (0.514)$ &                         &                        \\
	$R-i'$        & $0.454\pm0.345\ (0.461)$ & $0.278\pm0.239\ (0.289)$ & $0.349\pm0.294\ (0.303)$\\
	              & $0.531\pm0.279\ (0.542)$ &                         &                        \\
	$i'-z'$       & $0.314\pm0.512\ (0.378)$ & $0.271\pm0.194\ (0.275)$ & $0.243\pm0.198\ (0.241)$\\
	              & $0.483\pm0.253\ (0.489)$ &                         &                        \\
	$i'$-band magnitude & $24.17\pm1.82\ (24.11)$  & $22.23\pm0.81\ (22.04) $ & $22.04\pm0.98\ (21.85)$\\
	                    & $22.54\pm0.94\ (22.70)$  &                         &                        \\
	HR2                 & $-0.335\pm0.501\ (-0.506)$ & $-0.599\pm0.209\ (-0.637)$ & -\\
	                    & $-0.341\pm0.465\ (-0.501)$ &                         &                        \\
	$\log(f_{\rm{0.5-2.0keV}})$  & $-14.73\pm0.46\ (-14.68)$ & $-14.45\pm0.39\ (-14.44)$ & -\\
                                     & $-14.66\pm0.44\ (-14.63)$ &                         &                        \\
	$\log(f_{\rm{2.0-10.0keV}})$ & $-14.15\pm0.37\ (-14.13)$ & $-14.20\pm0.47\ (-14.17)$ & -\\
	                             & $-14.20\pm0.47\ (-14.17)$ &                         &                        \\
	$\log(f_{X\rm{0.5-2.0keV}}/f_{i'})$  & $0.400\pm0.745\ (0.393)$ & $-0.076\pm0.408\ (-0.054)$ & -\\
	                                     & $-0.149\pm0.500\ (-0.147)$ &                         &             \\
	$\log(f_{X\rm{2.0-10.0keV}}/f_{i'})$ & $0.998\pm0.810\ (1.005)$ & $0.171\pm0.491\ (0.261)$ & -\\
	                                     & $0.368\pm0.554\ (0.444)$ &                         &                  \\
	variable component $i'_{\rm{vari}}$& - & $23.90\pm0.82\ (24.03)$ & $24.69\pm0.84\ (24.96)$\\
	$\log(f_{i',{\rm{vari}}}/f_{i'})$  & - & $-0.666\pm0.305\ (-0.608)$ & $-1.059\pm0.512\ (-1.069)$\\
\enddata
\tablecomments{For the XA sample, upper rows are calculated using all the sample 
while lower rows are calculated using the objects with $i'<23.9$ mag.}
\label{tab:varixraycompare2}
\end{deluxetable}

\begin{deluxetable}{cc}
\tablewidth{0pt}
\tablecaption{K-S Test Probabilities of Optical Colors\label{tab:varixraycompare5}}
\tablehead{
\colhead{sample1} &
\colhead{LE-VA}\\
\colhead{sample2} &
\colhead{HE-VA}
}
\startdata
	$B-V$               & $3.32\rm{e}-03$ \\
	$V-R$               & $9.80\rm{e}-04$ \\
	$R-i'$              & $2.28\rm{e}-02$ \\
	$i'-z'$             & $3.96\rm{e}-01$ \\
\enddata
\tablecomments{K-S test probabilities between LE-VA and HE-VA objects for the optical colors.}
\label{tab:varixraycompare5}
\end{deluxetable}
\begin{deluxetable}{ccccc}
\tabletypesize{\scriptsize}
\tablewidth{0pt}
\tablecaption{Averages (Medians) and Standard Deviations of Optical Colors\label{tab:varixraycompare4}}
\tablehead{
\colhead{parameter} &
\colhead{XVA} &
\colhead{VA} &
\colhead{LE-VA} &
\colhead{HE-VA}
}
\startdata
	$B-V$                       & $0.360\pm0.438\ (0.357)$\tablenotemark{a} & $0.535\pm0.453\ (0.499)$\tablenotemark{a} & $0.734\pm0.383\ (0.683)$ & $0.401\pm0.448\ (0.400)$ \\
	$V-R$                       & $0.307\pm0.354\ (0.329)$\tablenotemark{a} & $0.431\pm0.419\ (0.379)$\tablenotemark{a} & $0.566\pm0.307\ (0.537)$ & $0.341\pm0.459\ (0.322)$ \\
	$R-i'$                      & $0.278\pm0.239\ (0.289)$\tablenotemark{a} & $0.349\pm0.294\ (0.303)$\tablenotemark{a} & $0.431\pm0.255\ (0.409)$ & $0.293\pm0.306\ (0.239)$ \\
	$i'-z'$                     & $0.271\pm0.194\ (0.275)$\tablenotemark{a} & $0.243\pm0.198\ (0.241)$\tablenotemark{a} & $0.253\pm0.168\ (0.269)$ & $0.237\pm0.215\ (0.230)$ \\
\enddata
\tablenotetext{a}{The same values as those in Table \ref{tab:varixraycompare2}.} 
\label{tab:varixraycompare4}
\end{deluxetable}

\end{document}